\UseRawInputEncoding
\documentclass[%
 twocolumn,
nofootinbib,
 amsfonts,amsmath,amssymb,
 aps,
prd,
floatfix,
showpacs,
longbibliography
]{revtex4-2}

\usepackage{graphicx}
\usepackage{dcolumn}
\usepackage{bm}
\usepackage{wasysym}
\usepackage{soul}
\usepackage{MnSymbol}
\usepackage{enumitem}
\usepackage{xfrac}
\usepackage[dvipsnames]{xcolor}
\definecolor{ReflexBlue}{rgb}{ .0902,.0902,.5882}
\usepackage[breaklinks]{hyperref}
\hypersetup{
    colorlinks=true,
    linkcolor=ReflexBlue,
    filecolor=magenta,      
    urlcolor=ReflexBlue,
    citecolor=ReflexBlue,
}
\usepackage{mathrsfs}

\begin{document}


\title{Backreaction from quantum fluxes at the Kerr inner horizon}

\author{Tyler McMaken}
 \email{Tyler.McMaken@colorado.edu}
\affiliation{%
 JILA and Department of Physics, University of Colorado, Boulder, Colorado 80309, USA
}%
\date{\today}

\begin{abstract}
Black holes modeled by the Kerr metric are not semiclassically self-consistent at or below the inner horizon. The renormalized stress-energy tensor (RSET) of a scalar quantum field in the Unruh state has been found to diverge at the Kerr inner horizon \cite{zil22b}, causing the geometry to backreact in a nontrivial way. In an effort to understand this backreaction, here the inner-horizon RSET is computed for the full physically relevant parameter space of black hole spins $a$ and polar angles $\theta$. Then, the backreaction is analyzed using a framework for the dynamical behavior of mass inflation from continued accretion. It is shown that the initial backreaction from the RSET does not evolve the spacetime toward any known regular or extremal configuration, but instead it brings the local interior geometry toward a chaotic, spacelike singularity, classically stable over astrophysical timescales.
\end{abstract}

\maketitle

\section{\label{sec:int}Introduction}

In general relativity, a rotating black hole is often modeled by the Kerr metric, an axisymmetric solution to Einstein's field equations in an empty spacetime. Within the Kerr black hole, an inner horizon marks the boundary of Cauchy predictability, and beyond it lies an observable singularity, closed timelike curves, and a wormhole to another external universe. But what if the spacetime is not empty? Is it possible for these structures to form in a real gravitational collapse? Put another way, what is the fate of the inner horizon and its traversability if a Kerr black hole is subject to perturbations?

In the case of classical perturbations, the inner horizon is subject to the mass inflation instability \cite{poi90}, an exponential divergence of the locally measured stress energy due to the presence of both ingoing and outgoing streams of radiation or matter accreted onto the black hole. The backreaction from this classical instability generically results in the geometry's collapse toward a strong, chaotic, oscillatory, spacelike singularity \cite{bra95,ham10,ham11a,ham11b,ham11c,ham17,mcm21,bar22}, though if the black hole remains isolated from all external influences aside from its own Price tail, the backreaction could result in a weak, null singularity \cite{ori92,ori99}.

In light of the classical divergence of stress energy near the inner horizon, one may wonder what role quantum effects play in the backreaction as the curvature approaches the Planck scale. If a field in a vacuum state ${|\psi\rangle}$ is canonically quantized over a curved spacetime, its stress-energy contribution to one-loop order can be renormalized and used as a source for semiclassical field equations \cite{mol62,ros63} (throughout, assume geometrized units where $G$=$c$=$\hbar$=1):
\begin{equation}\label{eq:EinsteinRSET}
    G_{\mu\nu}=8\pi\langle\psi|\hat{T}_{\mu\nu}|\psi\rangle_{\text{ren}}.
\end{equation}

\begin{figure}[t]
\centering
\includegraphics[width=0.95\columnwidth]{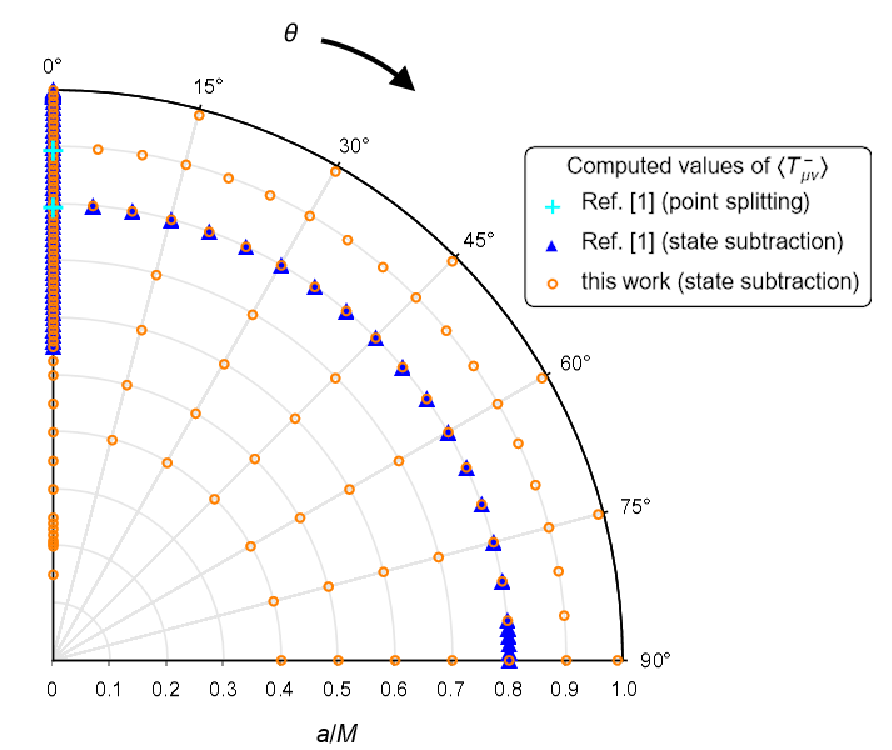}
\caption{Parameter space of Kerr spin-to-mass ratios ${a/M}$ and polar angles $\theta$ for which the dominant components of the RSET ${\langle T_{\mu\nu}^-\rangle}$ have been calculated for a scalar field at the inner horizon. Cyan plus signs indicate values calculated in Ref.~\cite{zil22b} via point-splitting, solid blue triangles indicate values calculated in Ref.~\cite{zil22b} via state subtraction, and hollow orange circles show the values calculated in the present study via state subtraction.\label{fig:parameters}}
\end{figure}

The computation of this renormalized stress-energy tensor (RSET) is a difficult task for physically realistic vacuum states in even the most symmetric spacetimes. But in recent years, novel computational methods have led to a resurgence of interest, and\textemdash most relevant to the present analysis\textemdash Ref.~\cite{zil22b} made use of state subtraction to calculate the mode-summed RSET of a scalar field at the inner horizon of a four-dimensional (4D) Kerr black hole (see \cite{zil22a} for a thorough derivation and Fig.~\ref{fig:parameters} for the parameter space covered by the study). The main conclusion of the study was that the ingoing double-null flux component of the RSET (the $vv$-component, in Eddington-Finkelstein coordinates) is generically nonvanishing, implying that the semiclassical stress energy will diverge when cast into coordinates that are regular across the inner horizon (as seen, for example, in the local frame of an infalling, outgoing\footnote{Infalling (outfalling) here means that the radial component of the observer's 4-velocity satisfies ${\dot{r}<0}$ (${\dot{r}>0}$), while ingoing (outgoing) refers to a left-moving (right-moving) observer on a traditional Penrose diagram. Ingoing observers are not considered here, since the Unruh state differs substantially from the more realistic Minkowski in-state along that portion of the inner horizon.} observer reaching the inner horizon). The implications are much stronger than in the classical case\textemdash even in the complete absence of external matter or radiation tails, the mere presence of an initially empty quantum field is enough to interrupt the gravitational collapse toward the full Kerr spacetime once the inner horizon is reached.

The divergence of the RSET at the inner horizon implies that the full geometry of the Kerr metric is not semiclassically self-consistent and must be substantially modified by the field's backreaction. Reference~\cite{zil22b} suggested that the backreaction should depend crucially on the signs of the RSET's null flux components, based on prior results from the spherically symmetric case \cite{zil20}: if the RSET's $vv$-component at the inner horizon is positive (negative), an observer approaching the inner horizon would experience an abrupt contraction (expansion). Since this component was generically found to change signs for different latitudes along the inner horizon and for different black hole spins, the result to first order is a chaotic inner-horizon singularity with local patches of abrupt contraction or expansion.

A more comprehensive analysis of the semiclassical backreaction in rotating black holes was recently carried out in the context of the Kerr\textendash de Sitter spacetime, confirming the postulations of Ref.~\cite{zil22b} that the RSET's $vv$-component dominates the evolution \cite{kle24a}. But if the RSET's $vv$-component changes sign across different latitudes, this would imply that some portions of the inner horizon would necessarily remain unscathed by any quantum null fluxes. Could a finely tuned observer evade the semiclassical singularity at the inner horizon? Not quite\textemdash Ref.~\cite{kle24a} found that even in these regions, a subdominant divergence in the RSET's $v\varphi$-component causes the local patch of geometry to experience a diverging amount of relative twist.

The goal of the present study is to provide independent validation of the prior studies' numerical work, extend the parameter space of known RSET values (see Fig.~\ref{fig:parameters}), and understand as much as can be possibly gleaned about the Kerr metric's semiclassical evolution out of equilibrium. There is a hope that the semiclassical gravity framework may be sufficient to describe the (meta)stable end point of evolution from a gravitational collapse toward a black hole \cite{dif24}\textemdash though the black hole may evaporate on supercosmological timescales, the mass-inflation and semiclassical instabilities operate on light-crossing timescales and may push the black hole toward a stable configuration with extremal horizons, a regular core, a wormhole throat, or no horizon at all (such as a gravastar). However, here evidence is given against the formation of such objects from Kerr-like initial conditions. Instead, an analysis of the backreaction using the mass-inflationary framework of Refs.~\cite{ham11a,ham11b,ham11c} suggests that the local geometry near the would-be inner horizon should evolve in a Belinskii-Khalatnikov-Lifschitz (BKL)-like fashion \cite{bel70} toward a strong, spacelike singularity. 

The derivation of the key details in the calculation of the RSET is presented in Sec.~\ref{sec:eff}, the numerical results of the calculations are presented in Sec.~\ref{sec:num}, and the analysis of the backreaction is given in Sec.~\ref{sec:bac}. In this analysis, it should be noted here that one cannot actually conclude definitively that a spacelike singularity is the generic outcome of the semiclassical instability near the inner horizon. The backreaction of the Kerr RSET is only valid while the geometry can still be well approximated by the vacuum Kerr metric, and as soon as the spacetime begins to evolve away from equilibrium, a new, nonvacuum RSET would need to be found. Most prior studies remain silent on the search for a path forward in our understanding other than an admission to the necessity of a full theory of quantum gravity. But here, another path is suggested: the local inner-horizon geometry under some mild assumptions becomes elegantly simple during the initial stages of mass inflation, and it thus may be possible in the future to compute the RSET in this simplified spacetime to determine the full semiclassical evolution of the near-inner-horizon geometry until the curvature reaches the Planck scale.

\section{\label{sec:eff}Renormalization of the stress-energy tensor}

The Kerr RSET at the inner horizon is here computed using the state subtraction method employed in Refs.~\cite{zil22a,zil22b}. The key details are outlined in what follows.

The Kerr line element for a black hole with mass $M$ and angular momentum ${J\equiv aM}$ can be written in Boyer-Lindquist coordinates as
\begin{align}\label{eq:kerr}
    ds^2=\rho^2\bigg(&\frac{dr^2}{(r^2+a^2)^2\Delta_r}+\frac{\sin^2\!\theta\ d\theta^2}{\Delta_\theta}\nonumber\\
    &+\frac{\Delta_\theta(d\varphi-\Omega_r dt)^2-\Delta_r(dt-\Omega_\theta d\varphi)^2}{(1-\Omega_r\Omega_\theta)^2}\bigg),
\end{align}
with the conformal factor $\rho^2\equiv r^2+a^2\cos^2\!\theta$, the horizon function $\Delta_r\equiv(r^2+a^2-2Mr)/(r^2+a^2)^2$ (with two roots at the outer and inner horizon radii $r_\pm\equiv M\pm\sqrt{M^2-a^2}$), the polar function $\Delta_\theta\equiv\sin^2\!\theta$, the angular velocity $\Omega_r\equiv a/(r^2+a^2)$ of the principal frame through the coordinates, and the specific angular momentum $\Omega_\theta\equiv a\sin^2\!\theta$ of principal null congruence photons.

If a massless scalar test field $\Phi$ is minimally coupled to the Einstein-Hilbert action and canonically quantized, the resulting Klein-Gordon wave equation \mbox{$\square$}$\Phi=0$ is separable \cite{car68,teu72} and lends itself to the field mode decomposition
\begin{equation}
    \Phi_{\omega\ell m}(t,r,\theta,\varphi)=\text{e}^{im\varphi-i\omega t}S_{\ell m}^{\omega}(\theta)R_{\omega\ell m}(r),
\end{equation}
where ${S_{\ell m}^{\omega}(\theta)}$ are prolate spheroidal harmonics, normalized on the two-sphere according to the Meixner-Sch{\"a}fke scheme (such a normalization adds some extra factors to the calculation but is the scheme of choice for the functions provided in \textit{Mathematica}) \cite{mei54}.

The field $\Phi$ is then equipped with an Unruh-type vacuum state ${|U\rangle}$ \cite{unr76,kle23} to mimic the effects of a physically realistic gravitational collapse, with positive frequencies along the past null boundaries defined with respect to the proper time of an infalling observer asymptotically approaching those radii \cite{mcm24}.

In order to calculate the Unruh-state RSET at the inner horizon, Ref.~\cite{zil22b} constructed a bare mode sum composed of the appropriate differential operator acting on the Hadamard two-point function. Since this sum is formally divergent, the summand was then subtracted from that of another bare mode sum equipped with a vacuum state known to lead to a vanishing RSET at the inner horizon, in order to yield a finite RSET. This latter vacuum state, constructed from a time-reversed, negative-mass Kerr spacetime, may appear unconventional but nonetheless agrees with the Unruh-state RSET obtained from a ``comparison'' state subtraction \cite{kle24a} as well as from a traditional point-splitting approach \cite{zil22b} (at least for two different values of the black hole spin parameter on the polar axis; see Fig.~\ref{fig:parameters}).

Computing the two-point function at the inner horizon for a field initialized at the past null boundaries necessitates solving a relativistic scattering problem. Although the Unruh-state modes at the past horizon are not eigenmodes of the radial Teukolsky equation, they can nonetheless be Fourier-decomposed into an orthonormal set of eigenmodes $R^{\text{in}}_{\omega\ell m}$ (originating from past null infinity) and $R^{\text{up}}_{\omega\ell m}$ (originating from the past horizon), so that the two-point function and therefore the RSET at the inner horizon can be expressed in terms of numerically attainable 1D scattering coefficients \cite{zil22a}. These coefficients are calculated using the Mano-Suzuki-Takasugi (MST) method \cite{man96,man97}, using a version of the Black Hole Perturbation Toolkit (BHPT) \cite{bhp} adapted by the author; see Appendix A of Ref.~\cite{mcm24} for details. In the notation\footnote{The translation from the notation of MST, BHPT, and the present work to that of Ref.~\cite{zil22b} can be accomplished by setting $\rho^{\text{up}}_{\omega\ell m}={(1-a^2)^{-2i\omega_+(1-a^2)^{-1/2}}C^{\text{ref}}_{\text{ext}}/C^{\text{inc}}_{\text{ext}}}$ and $B_{\omega\ell m}={(1-a^2)^{im/a}(r_-/r_+)^{1/2}B^{\text{trans}}_{\text{int}}/B^{\text{trans}}_{\text{ext}}}$, with the remaining coefficients obtained from the Wronskian conditions of Ref.~\cite{mcm24}.} of MST, these scattering coefficients satisfy the asymptotic conditions
\begin{subequations}
\begin{align}\label{eq:Rin_scattering}
    R^{\text{in}}_{\omega\ell m}&\to
    \begin{cases}
        B^{\text{ref}}_{\text{ext}}r^{-1}\text{e}^{i\omega r^*}+B^{\text{inc}}_{\text{ext}}r^{-1}\text{e}^{-i\omega r^*},&r\to\infty\\
        B^{\text{trans}}_{\text{ext}}\text{e}^{-i\omega_+r^*},&r\to r_+\\
        B^{\text{ref}}_{\text{int}}\ \text{e}^{i\omega_-r^*}+B^{\text{trans}}_{\text{int}}\text{e}^{-i\omega_-r^*},&r\to r_-
    \end{cases},\\
\label{eq:Rup_scattering}
    R^{\text{up}}_{\omega\ell m}&\to
    \begin{cases}
        C^{\text{trans}}_{\text{ext}}r^{-1}\text{e}^{i\omega r^*},&r\to\infty\\
        C^{\text{inc}}_{\text{ext}}\ \text{e}^{i\omega_+r^*}+C^{\text{ref}}_{\text{ext}}\text{e}^{-i\omega_+r^*},&r\to r_+\\
        C^{\text{trans}}_{\text{int}}\ \text{e}^{i\omega_-r^*}+C^{\text{ref}}_{\text{int}}\text{e}^{-i\omega_-r^*},&r\to r_-
    \end{cases},
\end{align}
\end{subequations}
where ${\omega_\pm\equiv\omega-m\Omega_r(r_\pm)}$, and the tortoise coordinate $r^*$ is chosen to be
\begin{equation}
    r^*\equiv r+\frac{1}{2\kappa_+}\ln\left|\frac{r-r_+}{2M}\right|+\frac{1}{2\kappa_-}\ln\left|\frac{r-r_-}{2M}\right|,
\end{equation}
with the surface gravity ${\kappa_\pm\equiv Mr_\pm\Delta_r'(r_\pm)}$ (where a prime denotes differentiation with respect to $r$), defined to be negative at the inner horizon.

In terms of these scattering coefficients, the state-subtracted mode sum yielding the RSET at the inner horizon can be written explicitly \cite{zil22a,kle24a}. In terms of the interior Eddington-Finkelstein coordinates ${u\equiv r^*-t}$ and ${v\equiv r^*+t}$, the three most relevant components of the RSET are the $vv$-, $v\varphi$-, and $uu$-components, which are given by the equations \cite{zil22b,kle24a}
\begin{subequations}
\begin{align}\label{eq:Tvv}
    \langle T_{vv}^-\rangle&=\sumint\left(E_{vv(\omega\ell m)}^{U-}-\coth\frac{\pi\omega_-}{-\kappa_-}\right),\\
\label{eq:Tvphi}
    \langle T_{v\varphi}^-\rangle&=\sumint\frac{-m}{\omega_-}\left(E_{vv(\omega\ell m)}^{U-}-\coth\frac{\pi\omega_-}{-\kappa_-}\right),\\
\label{eq:Tdiff}
    \langle T_{uu}^-\rangle-\langle T_{vv}^-\rangle&=\sumint\left(\coth\frac{\pi\omega_+}{\kappa_+}-1\right)\left(1-\left|\frac{C^{\text{ref}}_{\text{ext}}}{C^{\text{inc}}_{\text{ext}}}\right|^2\right).
\end{align}
\end{subequations}
Here and throughout, each RSET component is written as ${\langle T_{\mu\nu}^\pm\rangle\equiv\langle U|\hat{T}_{\mu\nu}|U\rangle_{\text{ren}}|_{r_\pm}}$, and the symbols $\sumint$ and ${E^{U-}_{vv(\omega\ell m)}}$ are defined by
\begin{align}\label{eq:sumint}
    \sumint&(X)\equiv\nonumber\\
    &\int_{0}^{\infty}\sum_{\ell=0}^{\infty}\sum_{m=-\ell}^{\ell}\frac{(2\ell+1)(\ell-m)!}{(\ell+m)!}\frac{\omega_-|S_{\ell m}^{\omega}(\theta)|^2}{32\pi^2Mr_-}(X)\ d\omega,
\end{align}
\begin{align}
    E_{vv(\omega\ell m)}^{U-}&=\nonumber\\
    \frac{r_-\omega_-}{r_+\omega_+}\Bigg[&\coth\frac{\pi\omega_+}{\kappa_+}\left(\left|\frac{B^{\text{ref}}_{\text{int}}}{B^{\text{trans}}_{\text{ext}}}\right|^2+\left|\frac{C^{\text{ref}}_{\text{ext}}}{C^{\text{inc}}_{\text{ext}}}\right|^2\left|\frac{B^{\text{trans}}_{\text{int}}}{B^{\text{trans}}_{\text{ext}}}\right|^2\right)\nonumber\\
    &+2\text{csch}\frac{\pi\omega_+}{\kappa_+}\Re\left(\frac{C^{\text{ref}}_{\text{ext}}B^{\text{ref}}_{\text{int}}B^{\text{trans}}_{\text{int}}}{C^{\text{inc}}_{\text{ext}}(B^{\text{trans}}_{\text{ext}})^2}\right)\nonumber\\
    &+\left(1-\left|\frac{C^{\text{ref}}_{\text{ext}}}{C^{\text{inc}}_{\text{ext}}}\right|^2\right)\left|\frac{B^{\text{trans}}_{\text{int}}}{B^{\text{trans}}_{\text{ext}}}\right|^2\Bigg].
\end{align}
Thus, the RSET at the Kerr inner horizon can be calculated in a straightforward manner by numerically solving the radial and angular Teukolsky equations and applying their eigenmode solutions to Eqs.~(\ref{eq:Tvv})\textendash(\ref{eq:Tdiff}).

\section{\label{sec:num}Numerical results}

\begin{figure}[b]
\centering
\includegraphics[width=0.88\columnwidth]{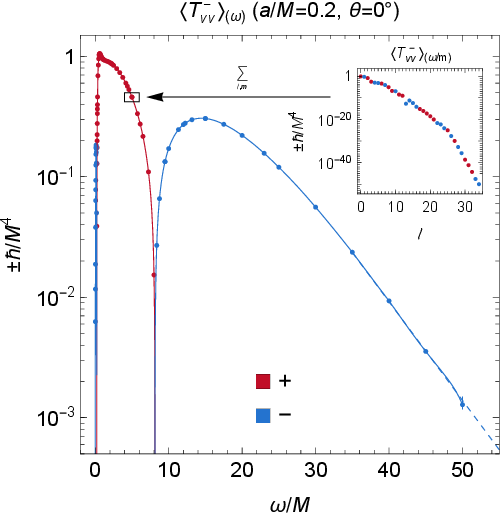}
\caption{Spectrum of ${\langle T_{vv}^-\rangle}$ at the north or south pole (${\theta=0^\circ}$) for a Kerr black hole with ${a/M=0.2}$. The sum over the $\ell$- and $m$-modes in the inset panel yields each point in the main panel (note that only the ${m=0}$ modes contribute when ${\theta=0^\circ}$), and the integral over the $\omega$-modes in the main panel yields the total RSET ${\langle T_{vv}^-\rangle}$ (i.e., the point at ${a/M=0.2}$ in Fig.~\ref{fig:polarfluxes}). Positive (negative) values on the log plots are given by the red (blue) points. The sampling and integration is performed using the adaptive quadrature algorithm described in the text.\label{fig:spectrum}}
\end{figure}

In order to make the calculation of the RSET as efficient as possible, integration is performed in the present work using an adaptive quadrature algorithm, which has the advantages of both rapid convergence for preferably sparse domain sampling and fine control over error estimation. An example of the output of this scheme for a particular choice of parameters can be seen in Fig.~\ref{fig:spectrum}. The ultimate goal is to integrate and sum Eqs.~(\ref{eq:Tvv})\textendash(\ref{eq:Tdiff}) over all possible $\omega$-, $\ell$-, and $m$-modes, which for brevity are written with the notation 
\begin{equation}
    \langle T_{\mu\nu}^-\rangle\equiv\int_0^\infty\langle T_{\mu\nu}^-\rangle_{(\omega)} d\omega\equiv\int_0^\infty\sum_{\ell=0}^{\infty}\sum_{m=-\ell}^{\ell}\langle T_{\mu\nu}^-\rangle_{(\omega\ell m)}d\omega.
\end{equation}
Each point in the main panel of Fig.~\ref{fig:spectrum} corresponds to the integrand ${\langle T_{vv}^-\rangle_{(\omega)}}$ for a particular value of the frequency $\omega$, and each of these points is computed by summing over all possible $\ell$- and $m$-modes (see inset of Fig.~\ref{fig:spectrum}). These sums converge exponentially and usually require only modes with ${\ell\leq2}$ for the error threshold to be vastly dwarfed by the error in the $\omega$ integration, though in some cases like the one shown in the figure, a larger number of angular modes are needed. 

Then, the algorithm proceeds by (i) computing four ${\langle T_{\mu\nu}^-\rangle_{(\omega)}}$ points subdivided evenly in a closed frequency domain, (ii) interpolating between these points with a cubic B-spline, (iii) computing a fifth sample point in the center of the domain, (iv) calculating the difference between this sample point and the center point of the spline, and (v) if this difference lies above a set error threshold, dividing the domain evenly in half and repeating the algorithm from step (i) for each new domain. This approach leads to a speedup of up to a factor of 10 in the total computational runtime compared to a fixed linear integration method and additionally allows for more precise error control.

Adaptive quadrature approaches perform most poorly in regimes where the integrand cannot be well approximated by polynomial functions, which for the integrands of Eqs.~(\ref{eq:Tvv})\textendash(\ref{eq:Tdiff}) occurs in the exponential decay regime at high frequencies (notice in Fig.~\ref{fig:spectrum} the slight difference at ${\omega/M\sim50}$ between the interpolating function and the dashed exponential fit). In these cases, two options were tested, both yielding similar convergent results: either the splines can be computed over log-frequency space and the domain extended until convergence is reached, or enough points can be sampled so that the remaining portion of the integrand can be fitted to an exponentially decaying function and the integral extrapolated to infinity.

\begin{figure}[t]
\centering
\includegraphics[width=0.88\columnwidth]{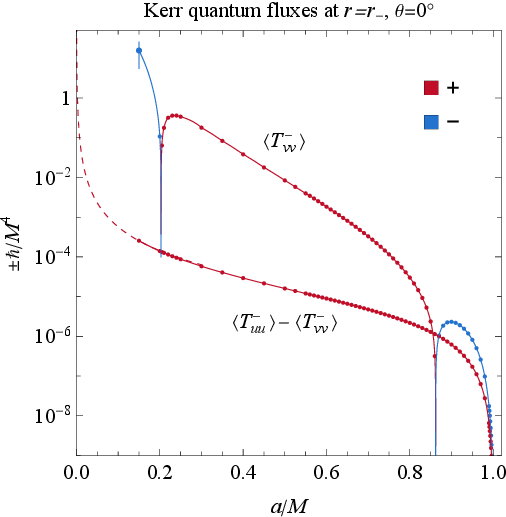}
\caption{Computed values of the $vv$- and $uu$-components of the RSET at the Kerr inner horizon along the axis of rotation for different values of the black hole spin-to-mass ratio ${a/M}$. Positive (negative) values on the log plot are given by the red (blue) points. Solid curves interpolate between the numerically computed points, and the dashed curve shows a 1/$a^2$ fit at low spins. \label{fig:polarfluxes}}
\end{figure}

\begin{figure}[t]
\centering
\includegraphics[width=0.9\columnwidth]{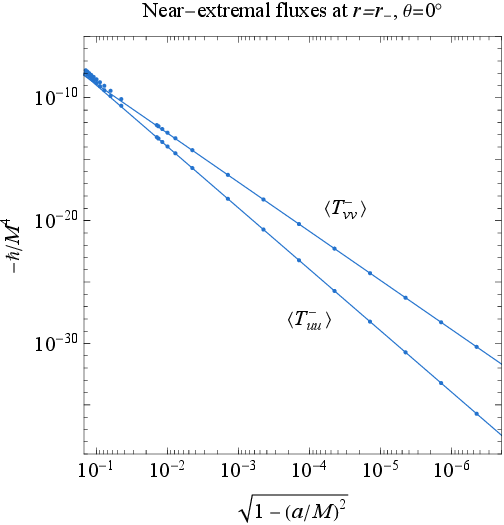}
\caption{Same setup as Fig.~\ref{fig:polarfluxes} but for black hole spins near extremality, with the near-zero parameter ${\epsilon\equiv\sqrt{1-(a/M)^2}}$. The two lines show the asymptotic behavior of the RSET's double-null components near extremality, given by the analytic expressions ${\langle T_{vv}^-\rangle\to\epsilon^4/(7680\pi^2)}$ and ${\langle T_{uu}^-\rangle\to\epsilon^5/(960\pi^2)}$.\label{fig:extremalpolarfluxes}}
\end{figure}

Two main sources of error are accounted for in the numerical calculations of the RSET performed here. The first is truncation error, which is minimized in the $\ell$- and $m$-sums by cutting off the sum only when the next term returns zero with the specified numerical precision, and which is controlled in the $\omega$-integrals by the degree of confidence in the exponential decay fit that is integrated to infinity. The second source of error is the global discretization error from the numerical integration scheme, which is controlled by specifying an error bound in the algorithm described above. Accounting for both of these sources of uncertainty, the points in Figs.~\ref{fig:spectrum}\textendash\ref{fig:extremalpolarfluxes} are computed to a high enough precision that their error bars are smaller than the points themselves in all but a couple of edge cases.

First, consider the inner-horizon RSET at the north or south pole, where ${\theta=0^\circ}$. There, only the modes with ${m=0}$ contribute to the RSET, so that ${\langle T_{v\varphi}^-\rangle}$ vanishes and the calculation of the other null components simplifies considerably. The two double-null flux components of the RSET are plotted as a function of the black hole's angular momentum $a$ in Figs.~\ref{fig:polarfluxes} and \ref{fig:extremalpolarfluxes}. Figure~\ref{fig:polarfluxes} shows computed values for spins from ${a/M=0.15}$ to ${a/M\approx0.997}$, yielding excellent quantitative agreement with Fig.~2 of Ref.~\cite{zil22b} and qualitative agreement with the Kerr\textendash de Sitter case in Fig.~4 of Ref.~\cite{kle24a} (both of which only reach a minimum of ${a/M=0.55}$). The $vv$-component of the RSET is negative and vanishingly small at high spins, and as the rotation of the black hole slows, this component increases until reaching zero at ${a/M\approx0.862}$ and continuing to increase exponentially in ${a/M}$. However, at slow enough spins, ${\langle T_{vv}^-\rangle}$ once again changes signs.

Both ${\langle T_{vv}^-\rangle}$ and ${\langle T_{uu}^-\rangle}$ are expected to diverge as ${a\to0}$, since in that limit the inner horizon coincides with the ${r=0}$ singularity. In particular, the quantity ${\langle T_{uu}^-\rangle-\langle T_{vv}^-\rangle}$, related to the Hawking outflux per surface area ${4\pi r_-^2}$ \cite{zil20}, should diverge as $a^{-2}$ (compare the dashed curve in Fig.~\ref{fig:polarfluxes}) owing to the factor of $r_-$ in the denominator of Eq.~(\ref{eq:sumint}). An even stronger divergence is expected to be present in ${\langle T_{vv}^-\rangle}$ as ${a\to0}$, since the scattering coefficients ${|B^{\text{ref}}_{\text{int}}|^2}$ and ${|B^{\text{trans}}_{\text{int}}|^2}$ from Eq.~(\ref{eq:Rin_scattering}) both diverge as $a^{-2}$. However, the exact nature of the low-spin divergence of ${\langle T_{vv}^-\rangle}$ is not easily found, since the spectrum includes nontrivial contributions from successively higher frequencies $\omega$ as the spin $a$ decreases, so that for ${a/M<0.15}$, ${\langle T_{vv}^-\rangle}$ could be either positive or negative (or even vanish in a highly fine-tuned case).

While the double-null components of the RSET are expected to diverge at the inner horizon as ${a/M\to0}$, they will vanish as ${a/M\to1}$, as shown in Fig.~\ref{fig:extremalpolarfluxes} (compare Fig.~3 in the Supplemental Material of Ref.~\cite{zil22b}). In particular, through an analysis of the asymptotic behavior of the near-extremal scattering coefficients, it can be shown that in terms of a near-extremal spin parameter ${\epsilon\equiv\sqrt{1-(a/M)^2}}$, the $vv$-component of the RSET will vanish as ${\langle T_{vv}^-\rangle\to\epsilon^4/(7680\pi^2)}$, while the $uu$-component will vanish as ${\langle T_{uu}^-\rangle\to\epsilon^5/(960\pi^2)}$.

Off the axis of rotation, the computation of the inner-horizon RSET is less clean. Even in the near-extremal limit, while the on-axis scattering coefficients are dominated by low-frequency ${\ell=0}$ modes that create a simple negative, exponentially decaying spectrum, the off-axis spectrum generally contains both positive and negative peaks at low frequencies. Nevertheless, in the limit as ${a/M\to1}$ for all polar angles $\theta$, all three components of the RSET analyzed here vanish.

\begin{figure}[t]
\centering
\includegraphics[trim={1.5cm 0 0 0},width=1.09\columnwidth]{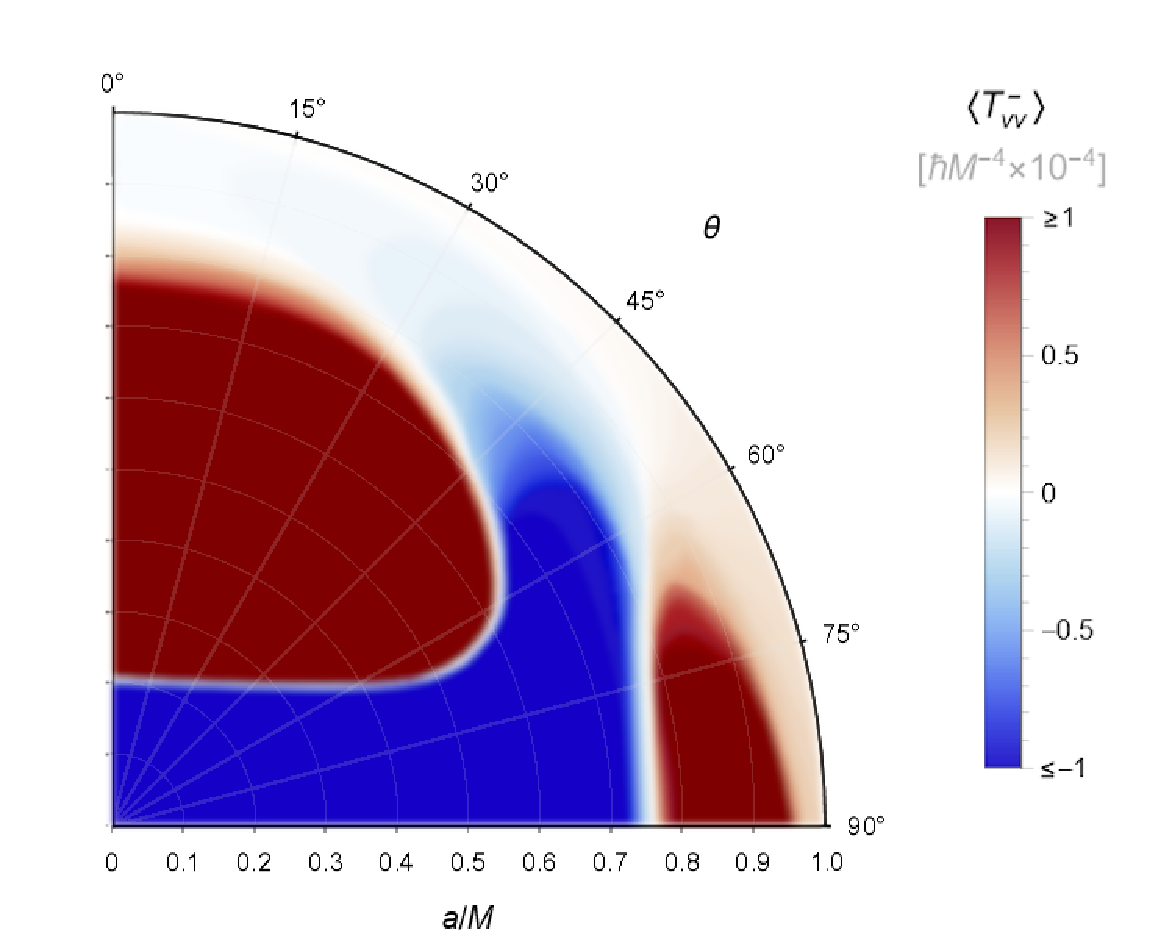}
\caption{Parameter space for the $vv$-component of the Kerr RSET at the inner horizon as a function of both the black hole spin-to-mass ratio ${a/M}$ and the polar angle $\theta$. The points at the locations given in Fig.~\ref{fig:parameters} were computed explicitly using the techniques described in the text, and the remaining portion of the parameter space was filled in via interpolation.\label{fig:Tvv_atheta}}
\end{figure}

The behavior of ${\langle T_{vv}^-\rangle}$ as a function of both spin $a$ and polar angle $\theta$ is shown in Fig.~\ref{fig:Tvv_atheta}. As can be seen, while the sign of ${\langle T_{vv}^-\rangle}$ changes as the spin of the black hole is varied, even for a single black hole with a given spin, ${\langle T_{vv}^-\rangle}$ will necessarily change signs at different latitudes along the inner horizon. For ${a/M\gtrsim0.862}$, ${\langle T_{vv}^-\rangle}$ is negative near the pole and reaches zero exactly once between the pole and the equator. For smaller spins, ${\langle T_{vv}^-\rangle}$ may change signs twice, though it should be noted that the behavior of the RSET for large $\theta$ and very small spins in Fig.~\ref{fig:Tvv_atheta} is only an extrapolated estimation.

Though not plotted here, the $v\varphi$-component of the RSET has also been computed and is generically nonzero throughout the 2D parameter space, except for at least one 1D zero-valued contour (just as ${\langle T^-_{vv}\rangle}$ in Fig.~\ref{fig:Tvv_atheta} contains two\footnote{Or three, if you count the $\theta$-parametrized contour at ${a/M=1}$, though in this work the extremal case is not treated.} zero-valued contours given by the white regions). Though the divergence associated with ${\langle T_{v\varphi}^-\rangle}$ is subdominant compared to ${\langle T_{vv}^-\rangle}$, the presence of a sign flip in ${\langle T_{vv}^-\rangle}$ implies that ${\langle T_{v\varphi}^-\rangle}$ will dominate the backreaction for at least one latitude, leading to a divergent twisting of the geometry separating regions of local expansion and contraction \cite{kle24a}. 

In conclusion, the renormalized stress energy of a scalar quantum field in the Unruh state has been calculated for most of the physically relevant parameter space in the Kerr spacetime at the inner horizon. The results indicate that in a locally inertial frame reaching the inner horizon, the field's flux generically diverges, and thus that the Kerr metric is not semiclassically self-consistent at or beyond the inner horizon. To understand how the geometry backreacts to this semiclassical instability, one must analyze the field equations, as shown in the next section.

\section{\label{sec:bac}Backreaction}

Analyses of semiclassical backreactions carry with them intricate subtleties. Ideally, one would like to find a solution to the M\o ller-Rosenfeld semiclassical field equations, Eq.~(\ref{eq:EinsteinRSET}), to see how the geometry and the quantum field coevolve over time. However, as can be gleaned from the calculations of the previous sections, the RSET is difficult to calculate and currently is only known for a select handful of highly symmetric vacuum spacetimes. Thus, the present calculations will apply only to a \emph{weak backreaction domain}, where one is still far enough from the inner horizon that the geometry can be well approximated by the Kerr metric, but close enough that the semiclassical backreaction, of order $(\text{e}^{\kappa_-v}\ m_P/M)^2$ (where $m_P$ denotes the Planck mass), is not negligible.

\subsection{Spherical symmetry}
The conclusions of any backreaction analysis are limited by the choice of assumptions about how the metric in question should be generalized. In the case of spherical symmetry (when ${a=0}$), such generalizations can be made comprehensively. If double-null coordinates $u$ and $v$ are gauge-fixed to match those of the vacuum RSET ${\langle T^-_{\mu\nu}\rangle}$, then the remaining two functional degrees of freedom in the line element can be written as $-\text{e}^{\sigma(u,v)}dudv+r(u,v)^2d\Omega^2$. The resulting field equations imply that near the inner horizon,
\begin{equation}
    \partial_vr\approx\frac{4\pi r_-}{\kappa_-}\langle T^-_{vv}\rangle+\mathcal{O}(\text{e}^{-\kappa_-v}).
\end{equation}
Recalling that the inner-horizon surface gravity $\kappa_-$ is defined here to be negative, one would then conclude that an infalling observer near the inner horizon would experience an abrupt contraction (expansion) in the geometry's area element when ${\langle T^-_{vv}\rangle}$ is positive (negative) \cite{zil20}.

However, if instead the radial coordinate $r$ is gauge-fixed to match that of the vacuum RSET and the remaining two metric degrees of freedom are encoded by\textemdash for example\textemdash the positions of the black hole's inner and outer horizons, the same field equations then imply dynamical behavior in the horizon structure rather than in the local area element. In particular, the outer horizon will evaporate inward slowly over time, as per Hawking's famous result (as long as ${\langle T^+_{vv}\rangle<0}$, which is always true in the 2D case) \cite{haw74,dav76,bar21}, while the inner horizon will evaporate rapidly outward (inward) when ${\langle T^-_{vv}\rangle}$ is negative (positive). The inner horizon in this case will move extremely quickly, at a timescale on the order of the black hole's light-crossing time \cite{bar21}.

While the two options described above for the spherically symmetric case might seem distinct or even at odds with one another, they are actually completely compatible. Extrapolating and reinterpreting those results as statements about the dominant contributions to global, long-term dynamics, ``which is irresistible but not allowed'' \cite{bal93}, would lead to the conclusion that black holes with ${\langle T^-_{vv}\rangle>0}$ have interiors that contract until a spacelike singularity is formed at ${r=0}$, while black holes with ${\langle T^-_{vv}\rangle<0}$ would explode from the inside out to form extremal or horizonless objects. However, since the RSET calculated here only remains valid in the weak backreaction domain, absolutely no conclusions can be made about long-term dynamics until the RSET is computed for the more general form of the spacetime.

\subsection{Axisymmetry: Initial tendencies}\label{subsec:axi_init}
Now, consider the case of black holes with nonzero rotation. Here it is clear that the outcome must be more complicated than in the spherically symmetric case, since the local interior geometry will no longer uniformly contract or expand across the entire inner horizon; instead, ${\langle T^-_{vv}\rangle}$ is both positive and negative across different latitudes of the same black hole (see Fig.~\ref{fig:Tvv_atheta}). An excellent backreaction analysis for the case of Kerr\textendash de Sitter black holes was recently \cite{kle24a} carried out by decomposing the spacetime into a set of double-null hypersurfaces gauge-fixed\footnote{The fixing comes in the identification of the affine parameter $\lambda$, taken to be constant along each hypersurface, with the Kruskal coordinate $V_-\propto\text{e}^{-\kappa_-v}$.} to match the null coordinates in ${\langle T^-_{\mu\nu}\rangle}$. Along each hypersurface, the induced metric can be written in a completely general form as
\begin{equation}
    g_{AB}dx^Adx^B=\gamma^2\left[\alpha^{-2}d\theta^2+\alpha^2\left(d\varphi+\tau d\theta\right)^2\right],
\end{equation}
for the area element $\gamma^2$ and the arbitrary functions $\alpha$ and $\tau$ that physically relate to notions of shear and twisting of null geodesics, as the hypersurface is moved toward the would-be inner horizon at ${\lambda\to0}$, or equivalently, ${v\to\infty}$. The semiclassical Einstein equations and Raychaudhuri equations can then be analyzed to yield
\begin{subequations}
\begin{align}
    &\partial^2_\lambda\gamma=-\gamma\left[4\pi\langle T^-_{\lambda\lambda}\rangle+4(\partial_\lambda\ln\alpha)^2+\alpha^4(\partial_\lambda\tau)^2\right],\label{eq:kle24a}\\
    &\partial_\theta(\gamma^2\alpha^4\partial_\lambda\tau)=8\pi\gamma^2\langle T^-_{\lambda\varphi}\rangle+\partial_\lambda(\gamma^2\beta),\label{eq:kle24b}
\end{align}
\end{subequations}
where $\beta$ measures the amount of twisting in the $\varphi$ direction for light rays perpendicular to the surfaces of constant $\lambda$ as ${\lambda\to0}$.

From Eq.~(\ref{eq:kle24a}), it can be gleaned that for positive (negative) ${\langle T^-_{vv}\rangle}$, the divergent RSET component ${\langle T^-_{\lambda\lambda}\rangle\sim\lambda^{-2}}$ will also be positive (negative) and will cause $\gamma$ to contract (expand) as ${\lambda\to0}$. In the inevitable intermediate cases where $\gamma$ neither contracts nor expands, Eq.~(\ref{eq:kle24b}) then predicts that a nonzero $\langle T^-_{v\varphi}\rangle$ will cause an infinite local twisting. Thus, one reproduces the same behaviors predicted in the spherically symmetric case (i.e., the local blow-up or shrinking of the geometry near the inner horizon dictated by the sign of ${\langle T^-_{vv}\rangle}$), with the addendum that even in the cases where the geometry neither contracts nor expands, a subdominant component of the RSET will still generically cause the local geometry to diverge in a shearing manner.

Two comments concerning the above analysis are worth mentioning. First, while the backreaction is performed with sufficient generality to reproduce the effects predicted in the spherical case and predict additional rotational effects, the restriction of the metric to null hypersurface cuts of constant $v$ does not allow for a full global analysis of the spacetime's evolution and does not encompass the most general axisymmetric geometry possible (it, for example, washes out any dynamical information related to the outgoing $u$ coordinate and its coupling with the other degrees of freedom). But more importantly, the analysis does not invite any obvious path forward to understand how the spacetime evolves beyond these initial tendencies. Does the semiclassical inner-horizon instability remain confined to a (meta)stable interior singularity (or even a regular configuration), or will inflationary perturbations spread like a wildfire until they destroy the black hole from the inside out? To address these problems at least partially, consider the complementary analysis below.

Generalizations of the Kerr metric abound. Starting from the line element of Eq.~(\ref{eq:kerr}), one may, for example, promote the mass $M$ to an arbitrary function ${M(r,\theta)}$ \cite{fer23}, or allow dynamical behavior via ${M\to M(v)}$ and ${a\to a(v)}$ \cite{yan11}, or leave all the functions $\Delta_r(r)$, $\Delta_\theta(\theta)$, $\Omega_r(r)$, $\Omega_\theta(\theta)$, and ${\rho^2=\rho_r^2(r)+\rho_\theta^2(\theta)}$ arbitrary for full Hamilton-Jacobi separability \cite{car68}. Additional generalizations beyond the form of Eq.~(\ref{eq:kerr}) also exist; imposing Klein-Gordon and timelike Hamilton-Jacobi separability yields a 3-function class of metrics that reproduces a wide variety of regular and singular spacetime candidates \cite{bai23}, and imposing only asymptotic flatness and the preservation of the Carter constant leads to a 10-function class of general axisymmetric metrics \cite{pap18}. Amidst all these options, one thus desires a trade-off between generality and tractability in capturing the most important physical behaviors to be modeled and understood.

The properties of the semiclassical radiation produced from a quantum scalar field in the vacuum Kerr spacetime already seem to rule out a number of Kerr generalizations. For example, the 3-function class of metrics put forward in Ref.~\cite{bai23} claims to encompass a majority of the currently proposed stable end points of black hole evolution (aside from complete evaporation or gravastar-like objects); however, the Einstein tensor for these metrics always contains a vanishing $rt$-component, despite the fact that the semiclassical RSET calculated here possesses a nonvanishing (and in fact diverging) $rt$-component ${\langle T^-_{rt}\rangle}={\left(\langle T^-_{vv}\rangle-\langle T^-_{uu}\rangle\right)R^{-2}\Delta_r^{-1}}$. Presumably, such black hole solutions that are regular and instability-free cannot form from an initial Kerr-like collapse.

\subsection{Axisymmetry: Mass inflationary approach}\label{subsec:axi_infl}
To generalize the Kerr metric in order to allow for the dynamics anticipated from the vacuum RSET, the following assumptions will be made:
\begin{enumerate}[label=(\roman*)]
\item The spacetime is axisymmetric, motivated both by simplicity and by the preservation of this symmetry in the past Unruh state.
\item The spacetime is asymptotically flat at spatial infinity, as in the case of the original Kerr metric.
\item The black hole is initially isolated as in the Kerr vacuum geometry, and the quantum scalar field is initialized with an Unruh vacuum state mimicking the effects of a gravitational collapse sufficiently far in the past.
\item The spacetime maintains conformal Hamilton-Jacobi separability \cite{ham11b}\textemdash i.e., the equations of motion for massless particles are separable in spheroidal coordinates (but not necessarily for massive particles, as in the case of strict separability seen for Kerr).
\end{enumerate}

Unfortunately, none of these assumptions are actually true for astrophysically realistic black holes. The accretion flows observed by the GRAVITY instrument \cite{gra18} and the Event Horizon Telescope Collaboration \cite{eht22} feature asymmetries in $\varphi$, though in the stationary limit, rigidity theorems suggest that any nonaxisymmetric perturbations are part of the ``hair'' that a black hole will eventually shed \cite{haw72,hol23}. In regard to assumption (ii), the lack of asymptotic flatness in the Universe has been measured with great precision \cite{pla16}, but any effects from a cosmological horizon on the inner-horizon instability are expected to be negligible, both due to the similarity in the RSET calculations here to those of the Kerr\textendash de Sitter case \cite{kle24a}, and because the cosmological constant is currently measured to be $\Lambda\sim10^{-46}\ M_{\astrosun}^{-2}$ in geometrized units, much too small to have any practical influence. As for assumption (iii), black holes (and therefore near-inner-horizon geometries) form under chaotic, rapidly evolving conditions, and even after settling down, the most quiescent black holes are still nonvacuum and accrete more than enough radiation to trigger the mass inflation instability and destroy any vacuum inner horizon \cite{mcm21}. Additionally, physically relevant fields in the Standard Model are those with spin 1 (electromagnetic), 1/2 (fermionic), or 2 (gravitational), so using a massless scalar field with spin 0 is a simplification seen as a proxy for, e.g., a single photonic degree of freedom. Even regarding assumption (iv), which provides a more general notion than the strict separability that itself is justified by the observation of long-lived accretion disks (lack of separability implies chaotic, destabilizing particle orbits \cite{bai23}), it will be seen in what follows that conformal separability, while valid in the weak backreaction domain, eventually breaks down once the vierbein of Eq.~(\ref{eq:dyn_vierbein}) ceases to be diagonal.

Under assumptions (i)\textendash(iv), the spacetime's line element takes the form of Eq.~(\ref{eq:kerr}), except that the functions $\Delta_r(r)$, $\Delta_\theta(\theta)$, $\Omega_r(r)$, $\Omega_\theta(\theta)$, and $\rho(r,\theta,t)$ are left arbitrary. Assumption (ii) of asymptotic flatness, while not strictly required for any aspect of this analysis (any cosmological contribution to the near-inner-horizon geometry will be completely overwhelmed by the ignition of both classical and semiclassical streams \cite{ham11b}), will impose the additional conditions $\Delta_r\sim\Omega_r\sim r^{-2}$ and $\rho\sim r$.

The form of the line element of Eq.~(\ref{eq:kerr}) naturally encodes an orthonormal tetrad basis first written down by Carter \cite{car68}. This basis can be constructed in the Newman-Penrose formalism by performing a null rotation on the Kinnersley tetrad so that the resulting frame is symmetric with respect to the ingoing and outgoing principal null directions \cite{zna77}. The corresponding vierbein $e^{\hat{m}}_{\ \mu}$, defined through
\begin{equation}
    ds^2=g_{\mu\nu}dx^\mu dx^\nu=\eta_{\hat{m}\hat{n}}e^{\hat{m}}_{\ \mu}e^{\hat{n}}_{\ \nu}dx^\mu dx^\nu
\end{equation}
for the Minkowski metric tensor $\eta_{\hat{m}\hat{n}}$, can be written as the product \cite{ham17}
\begin{equation}\label{eq:vierbein_decomp}
    e_{\hat{m}\mu}=(e_{\text{dyn}})_{\hat{m}\kappa}\ (e_{\text{fix}})^{\kappa}_{\ \mu},
\end{equation}
where $(e_{\text{dyn}})_{\hat{m}\kappa}$ is a dynamical vierbein that is only a function of the variable $r$ (which is timelike in the interior), and $(e_{\text{fix}})^{\kappa}_{\ \mu}$ is a fixed vierbein whose elements should remain frozen at their inner-horizon values throughout the evolution of the instability induced by the semiclassical backreaction. In terms of Boyer-Lindquist coordinates, the fixed vierbein contains the basis vectors
\begin{subequations}\label{eq:fix_vierbein}
\begin{align}\label{eq:fix0}
    (e_{\text{fix}})^0&=\rho\ \partial_r,\\\label{eq:fix1}
    (e_{\text{fix}})^1&=\frac{\rho}{1-\Omega_r\Omega_\theta}\left(\partial_t-\Omega_\theta\partial_\varphi\right),\\\label{eq:fix2}
    (e_{\text{fix}})^2&=\rho\ \partial_\theta,\\\label{eq:fix3}
    (e_{\text{fix}})^3&=\frac{\rho\sqrt{\Delta_\theta}}{1-\Omega_r\Omega_\theta}\left(\partial_\varphi-\Omega_r\partial_t\right),
\end{align}
\end{subequations}
which serves to align the tetrad with the principal null directions of the black hole. The dynamical vierbein then is purely diagonal:
\begin{subequations}\label{eq:dyn_vierbein}
\begin{align}\label{eq:dyn0}
    (e_{\text{dyn}})_0&=\frac{\text{e}^{-5\xi/2}}{R^2\sqrt{|\Delta_r|}}\ \partial_0,\\\label{eq:dyn1}
    (e_{\text{dyn}})_1&=\text{e}^{\xi/2}\sqrt{|\Delta_r|}\ \partial_1,\\\label{eq:dyn2}
    (e_{\text{dyn}})_2&=\text{e}^{-\xi}\ \partial_2,\\\label{eq:dyn3}
    (e_{\text{dyn}})_3&=\text{e}^{-\xi}\ \partial_3,
\end{align}
\end{subequations}
where the redefinitions ${\rho(r,\theta,t)\to\text{e}^{-\xi(r)}\rho(r,\theta,t)}$ and $\Delta_r(r)\to\text{e}^{3\xi(r)}\Delta_r(r)$ have been made for an arbitrary radial function $\xi(r)$, which is brought to the dynamical vierbein so that the vierbein may undergo a conformal expansion or contraction under backreaction.

The Einstein tensor corresponding to this spacetime is far too intricate to be presented here in any complete, meaningful way. However, at least in the quasistationary limit (in particular, if the redefined conformal piece $\rho$ is written as $\rho(r,\theta,t)=\text{e}^{\tilde{v}t}\sqrt{r^2+a^2\cos^2\!\theta}$ for some vanishing small accretion parameter $\tilde{v}$), the field equations have already been analyzed in detail in Refs.~\cite{ham11a,ham11b,ham11c}. The resulting behavior provides classical justification for the decomposition of Eq.~(\ref{eq:vierbein_decomp}), since near the inner horizon, the functions $\Delta_\theta$, $\Omega_r$, and $\Omega_\theta$ remain approximately fixed at their Kerr values near the inner horizon, while the horizon function $\Delta_r$ and the inflationary function $\xi$ become the dominant contributors to the evolution of the Einstein tensor. The resulting tensor behaves at least initially like a perfect fluid with equal ingoing and outgoing streams, with dominant inflating tetrad-frame components $G_{00}\sim G_{11}\sim G_{01}\sim\Delta_r^{-1}$, where the off-diagonal component $G_{01}$ is suppressed by a factor of $\tilde{v}$. Classically, the evolution can then be continued to show that $\Delta_r$ eventually stalls out at some tiny value \cite{ham11b}, after which $\xi$ grows large and the spacetime undergoes a series of BKL-like bounces toward a strong, spacelike singularity \cite{ham17}.

How might semiclassical effects modify this picture? The tetrad-frame components of the RSET can be reexpressed in terms of the double-null Eddington-Finkelstein coordinates used throughout this paper:
\begin{subequations}\label{eq:Tmn}
\begin{align}
    \langle T_{00}\rangle&=\frac{\text{e}^{5\xi}}{\rho^2|\Delta_r|}\bigg[\langle T_{uu}\rangle+\langle T_{vv}\rangle+2\langle T_{uv}\rangle\bigg],\\
    \langle T_{11}\rangle&=\frac{\text{e}^{-\xi}}{\rho^2|\Delta_r|}\bigg[\langle T_{uu}\rangle+\langle T_{vv}\rangle-2\langle T_{uv}\rangle\nonumber\\
    &-2\Omega_r\Big(\langle T_{u\varphi}\rangle-\langle T_{v\varphi}\rangle\Big)+\Omega_r^2\langle T_{\varphi\varphi}\rangle\bigg],\\
    \langle T_{01}\rangle&=\frac{2\text{e}^{2\xi}}{\rho^2|\Delta_r|}\bigg[\langle T_{uu}\rangle-\langle T_{vv}\rangle-\Omega_r\Big(\langle T_{u\varphi}\rangle+\langle T_{v\varphi}\rangle\Big)\bigg],\\
    \langle T_{13}\rangle&=\frac{\text{e}^{\xi/2}}{\rho^2\sqrt{|\Delta_r|\Delta_\theta}}\bigg[\Omega_\theta\Big(\langle T_{uu}\rangle+\langle T_{vv}\rangle-2\langle T_{uv}\rangle\Big)\nonumber\\
    &+(1+\Omega_r\Omega_\theta)\Big(\langle T_{v\varphi}\rangle-\langle T_{u\varphi}\rangle\Big)+\Omega_r\langle T_{\varphi\varphi}\rangle\bigg],\\
    \langle T_{22}\rangle&=\frac{\text{e}^{2\xi}\Delta_\theta}{\rho^2\sin^2\!\theta}\langle T_{\theta\theta}\rangle,\\
    \langle T_{33}\rangle&=\frac{\text{e}^{2\xi}}{\rho^2\Delta_\theta}\bigg[\Omega_\theta^2\Big(\langle T_{uu}\rangle+\langle T_{vv}\rangle-2\langle T_{uv}\rangle\Big)\nonumber\\
    &+2\Omega_\theta\Big(\langle T_{v\varphi}\rangle-\langle T_{u\varphi}\rangle\Big)+\langle T_{\varphi\varphi}\rangle\bigg].
\end{align}
\end{subequations}
Far enough away from the inner horizon, all components of the RSET should be completely negligible, as they are suppressed by the Planck scale. Thus, everywhere outside the black hole as well as inside when $v$ is not large, a vacuum source should recover the standard Kerr solution with $\xi=0$ and $\Delta_r=(r^2+a^2-2Mr)/(r^2+a^2)^2$. However, in the weak backreaction domain, once the null components of the RSET cease to be vanishingly small, Equation~(\ref{eq:Tmn}) indicates that the radial and time components of the tetrad-frame Einstein tensor will begin to diverge as $\Delta_r^{-1}$ as one takes $\Delta_r\to0$.

Though the RSET components ${\langle T^-_{uv}\rangle}$, ${\langle T^-_{u\varphi}\rangle}$, and ${\langle T^-_{\varphi\varphi}\rangle}$ have not been computed explicitly, one can make the assumption that their contributions to Eq.~(\ref{eq:Tmn}) will be subdominant compared to that of the double-null components, and more importantly, that they will not contrive to cause any of the specific combinations in the equations above to cancel exactly. Even if they do, then the classical mass inflation phenomenon described in Refs.~\cite{ham11a,ham11b,ham11c} will take over the evolution, and a spacelike singularity will form. But based on the numerical results in Sec.~\ref{sec:num}, it is apparent that ${\langle T^-_{00}\rangle}$ and ${\langle T^-_{11}\rangle}$ are the dominant contributors to the evolution in the weak backreaction domain, while the off-diagonal components ${\langle T^-_{01}\rangle}$ are also important but initially much smaller than their diagonal counterparts (recall from Fig.~\ref{fig:polarfluxes} that the difference ${\langle T^-_{uu}\rangle}-{\langle T^-_{vv}\rangle}$ is almost always several orders of magnitude smaller than either individual component).

The Einstein tensor combination $G_{00}+G_{11}$ for the tetrad frame of Eq.~(\ref{eq:vierbein_decomp}) can be written as
\begin{align}\label{eq:G00+G11}
    G_{00}+G_{11}&=-\frac{2\Delta_r}{\rho^2}\bigg(\partial^2_{r_*}\xi+\left(\partial_{r_*}\xi\right)^2\nonumber\\
    &-2\partial_{r_*}\ln\left(\frac{\rho}{1-\Omega_r\Omega_\theta}\right)\partial_{r_*}\xi\bigg)+F_{rt},
\end{align}
where the tortoise coordinate $r_*$ is defined by $dr/dr_*=\text{e}^{3\xi}R^2\Delta_r$, and the function $F_{rt}$ encompasses subdominant terms related to the precise nature of any classical accretion contributing to nonstationary conformal dynamics.

The dominant term in Eq.~(\ref{eq:G00+G11}) is the one involving the second derivative of the inflationary exponent $\xi$ \cite{ham11b}. Equating this term to the semiclassical source from Eq.~(\ref{eq:Tmn}), which behaves as $\Delta_r^{-1}$ near the inner horizon, yields after integration the approximate solution
\begin{align}\label{eq:dxi}
    \partial_{r_*}\xi&\approx\frac{8\pi}{-\Delta_r}\Big(\langle T^-_{uu}\rangle+\langle T^-_{vv}\rangle\nonumber\\
    &-2\Omega_r(\langle T^-_{u\varphi}\rangle-\langle T^-_{v\varphi}\rangle)+\Omega_r^2\langle T^-_{\varphi\varphi}\rangle\Big),
\end{align}
as long as $\xi$ remains smaller than its derivatives. In the regime where the double-null components of the RSET dominate over the shearing components, Eq.~(\ref{eq:dxi}) thus predicts that when the sum ${\langle T^-_{uu}\rangle+\langle T^-_{vv}\rangle}$ is positive (negative), the spacetime's conformal factor $\text{e}^{-\xi}$ will abruptly contract (expand) as the inner horizon is approached at $\Delta_r\to0_-$. These initial tendencies align precisely with that of the spherically symmetric case, since ${\langle T^-_{uu}\rangle+\langle T^-_{vv}\rangle}$ is usually the same sign as ${\langle T^-_{vv}\rangle}$.

One can continue the evolution of the spacetime by examining the behavior of the dynamical vierbein of Eqs.~(\ref{eq:dyn0})\textendash(\ref{eq:dyn3}). Under the assumption that the fixed vierbein of Eqs.~(\ref{eq:fix0})\textendash(\ref{eq:fix3}) remains stable as the initial backreaction ignites the inflationary tendencies described above, the geometry will be governed solely by a diagonal, dynamical, homogeneous line element sourced by inflating streams of semiclassical matter. The classical counterpart of this analysis has already been carried out in detail in Ref.~\cite{mcm21}, yielding the so-called inflationary Kasner metric: as one might anticipate, the streams of matter are amplified in energy density as $\Delta_r$ plunges to zero and $\xi$ increases rapidly. Then, the spacetime undergoes a bounce as $\Delta_r$ freezes out at a small value and the $\text{e}^{\xi/2}$ term in Eq.~(\ref{eq:dyn1}) takes over and causes a radial expansion. The solution works for both positive and negative sources of stress energy, generically producing a series of inflating Kasner-like bounces toward a spacelike singularity.

Eventually, the approximations made in this section will break down, and the final evolution of the geometry near the spacelike singularity must be relinquished to a higher-order theory of quantum gravity. However, the results remain robust in the weak backreaction domain, and even in the presence of additional non-negligible stress-energy sources not accounted for in the analysis above, it has been shown \cite{ham11b} that the dominant double-null contributors to the geometry's initial inflation remain dominant to the next-highest order, as long as one still has $|\Delta_r|\ll\partial_{r_*}\xi$ or $\partial_{r_*}\xi\ll1$. If anything, the semiclassical contribution to the classical mass inflation instability will cause the conformally separable solution to break down faster than it otherwise would (due to the presence of shearing terms that act to rotate the dynamical vierbein away from its initial configuration). But numerical work in the classical case serves to indicate that even with such perturbations, the generic result should still be a chaotic, BKL-like collapse toward a spacelike singularity \cite{ham17}.

\section{\label{sec:dis}Discussion}

By now it is clear in the literature that quantum fields do not jive well with vacuum black hole spacetimes. If the Kerr metric (the axisymmetric solution to Einstein's equations for a rotating black hole in an empty spacetime) is immersed within a scalar quantum field, then even if that field begins in an empty vacuum state, the gravitational collapse leading to the formation of the black hole will cause a mixing of positive- and negative-frequency modes that leads to the spontaneous production of particles. Usually, this semiclassical flux of particles will be suppressed on the order of the Planck mass $m_P^2$ divided by the black hole's mass $M^2$, and the Kerr structure will remain intact, but close enough to the inner horizon, this study confirms the conclusion of previous studies \cite{zil22b,kle24a,mcm24} that semiclassical radiation will diverge at the Kerr inner horizon.

To understand how the Kerr geometry will react to the diverging quantum field near the inner horizon, it is natural to work in the framework of the semiclassical Einstein field equations [Eq.~(\ref{eq:EinsteinRSET})], wherein the spacetime's curvature is sourced by the vacuum expectation value of the renormalized stress-energy tensor (RSET) of the quantum field. The potential problems (mathematical, physical, and philosophical) associated with such an approach have been debated again and again over the years \cite{fey96,aha03,opp23,kle24b}, but it nonetheless remains true that the semiclassical approach is perfectly valid as an effective field theory of quantum gravity, as long as the RSET remains below Planck-scale energies (at $10^{19}$ GeV, which is already orders of magnitude above the grand unified scale, beyond the point where every atomic nucleus has been dissociated and the very notion of particles and interactions is called into question).

Here it is found that the RSET in double-null coordinates contains a nonzero component ${\langle U|\hat{T}_{vv}|U\rangle_{\text{ren}}}$ near the Kerr inner horizon, which, when converted to coordinates that are regular across that horizon, yields an exponential divergence in the quantum flux. Such behavior was also found in Refs.~\cite{zil22b,kle24a}; the additional contributions here were to compute more inner-horizon RSET components over the full parameter space ($a$, $\theta$) (Fig.~\ref{fig:Tvv_atheta}), find the asymptotic behavior at both high and low black hole spins (Figs.~\ref{fig:polarfluxes} and \ref{fig:extremalpolarfluxes}), and analyze the backreaction that the RSET elicits.

The most striking feature of the inner-horizon RSET is that, in contrast to the classical case (or even the semiclassical 2D case), the double-null components of the RSET can be either positive or negative at different points along the inner horizon. The semiclassical Einstein equations then suggest that as an initial tendency, the local geometry as one approaches the inner horizon will rapidly contract (expand) wherever ${\langle T^-_{vv}\rangle}$ is positive (negative).

In semiclassical backreaction analyses, the above statement is usually the end of the story. Reference~\cite{kle24a} takes it one step further to show that in the inevitable regions where the local geometry neither contracts nor expands, the $v\varphi$-component of the RSET will cause an initial tendency for the geometry to undergo an infinite twisting. However, the question of whether anything more can be ascertained from the RSET about the geometry's evolution and the black hole's final configuration is either left to speculation or completely ignored, and for good reason\textemdash as soon as the inner horizon is pushed even slightly away from its Kerr value, the Kerr vacuum RSET is no longer valid and a new RSET in a more general spacetime would need to be found to continue the evolution any further. Nonetheless, the goal of the current work is to provide a path forward to do exactly that, so that one may determine whether or not the semiclassical inner-horizon instability generically leads to a spacelike singularity.

The backreaction analysis of Sec.~\ref{subsec:axi_infl} employed the conformally separable framework of Refs.~\cite{ham11a,ham11b,ham11c,ham17} to determine how semiclassical fluxes can be understood together with the classical fluxes that lead to mass inflation near the inner horizon. One can immediately gather from the presence of a nonvanishing RSET component ${\langle T^-_{rt}\rangle}={\left(\langle T^-_{vv}\rangle-\langle T^-_{uu}\rangle\right)R^{-2}\Delta_r^{-1}}$ that the semiclassical backreaction drives the evolution away from many of the potential (meta)stable geometries proposed as potential end points of black hole evolution, like the regular ``eye of the storm'' geometry or the Kerr ``black-bounce'' geometry (see the end of Sec.~\ref{subsec:axi_init}). In contrast, the conformally separable model generically predicts the presence of a strong, spacelike singularity that is stable over astrophysical timescales.

In particular, in Sec.~\ref{subsec:axi_infl}, the metric is generalized to allow for all the functional degrees of freedom in the Kerr line element of Eq.~(\ref{eq:kerr}) to respond freely to the presence of semiclassical stress energy. The dominant evolution near the inner horizon comes from the horizon function $\Delta_r(r)$ and the conformal factor $\rho(r,\theta,t)$, and thus it becomes useful to analyze the spacetime in a tetrad frame rotated to align with the principal null directions so that the $\Delta_r$- and $\rho$-dependent portions of the vierbein are purely diagonal, Eqs.~(\ref{eq:dyn0})\textendash(\ref{eq:dyn3}).

In this tetrad frame, the RSET's dominant components are the diagonals ${\langle T^-_{00}\rangle}$ and ${\langle T^-_{11}\rangle}$ [Eq.~(\ref{eq:Tmn})], while the off-diagonal ${\langle T^-_{01}\rangle}$ also diverges as $\Delta_r^{-1}$ but is numerically seen to be usually several orders of magnitude smaller than the diagonal components. As such, one is justified in treating the initial semiclassical evolution with the inflationary Kasner metric \cite{mcm21}, which uses precisely the dynamical vierbein of Eqs.~(\ref{eq:dyn0})\textendash(\ref{eq:dyn3}) sourced by stress-energy components $T_{00}$ and $T_{11}$. Eventually (or even initially, in the latitudes of infinite quantum twisting), this solution will break down once the RSET's angular components become non-negligible, but even once the inflationary Kasner model breaks down, the full conformally separable model can still be continued (albeit with less symmetry), as it has been shown to be robust to a wide variety of stress-energy sources \cite{ham11b}.

The advantage of the conformally separable model at describing the inner-horizon backreaction is not only in its ability to encompass a wide range of behaviors related to the mass inflation phenomenon, but also in its potential to take the semiclassical evolution even deeper. It is almost miraculous that such a chaotic instability leads at least initially to a spacetime (the inflationary Kasner metric) that is so simple, elegant, and possesses such a high degree of symmetry. The RSET in this new homogeneous spacetime should be relatively straightforward to compute, with the only difficulty coming from mode-matching to the Kerr Unruh state (though the relevant mathematical details have already been worked out in Ref.~\cite{ham22}). In fact, the renormalized vacuum polarization $\langle\hat{\phi}^2\rangle_{\text{ren}}$ has already been calculated in the inflationary Kasner spacetime in an adiabatic vacuum state \cite{mcm22}, with the conclusion that the semiclassical contribution follows the same qualitative behavior as the classical stress energy.

It is time to take RSET calculations beyond vacuum black holes. From this work, it can be gleaned that the semiclassical inner-horizon instability causes the local Kerr spacetime to be filled with diverging streams of radiation, and while this work provides evidence that those streams should generically lead to an astrophysically stable, chaotic, spacelike singularity, their exact semiclassical evolution remains an open but tractable question.

\acknowledgments
I would like to thank Andrew J.~S.\ Hamilton, Adrian Ottewill, and Noa Zilberman for many discussions that helped shape the ideas of this work. This project is supported solely through teaching funds provided by the Department of Physics at the University of Colorado Boulder.

\bibliography{apsbib}

\end{document}